# Energetics, structure, and composition of nanoclusters in Oxide Dispersion Strengthened Fe-Cr alloys

M. Posselt[*], D. Murali

*Helmholtz-Zentrum Dresden-Rossendorf,*

*Institute of Ion Beam Physics and Materials Research,*

*P. O. Box 510119, D-01314 Dresden, Germany*

B. K. Panigrahi

*Indira Gandhi Centre for Atomic Research,*

*Materials Science Group,*

*Kalpakkam 603102, India*

**Abstract**

Extensive first-principle calculations on embedded clusters containing few O, Y, Ti, and Cr atoms as well as vacancies are performed to obtain interaction parameters to be applied in Metropolis Monte Carlo simulations, within the framework of a rigid lattice model. A novel description using both pair and triple parameters is shown to be more precise than the commonly used pair parameterization. Simulated annealing provides comprehensive data on the energetics, structure and stoichiometry of nm-size clusters at $T=0$. The results are fully consistent with the experimental finding of negligible coarsening and a high dispersion of the clusters, with the observation that the presence of Ti reduces the cluster size, and with the reported radiation tolerance of the clusters. In alloys without vacancies clusters show a planar structure, whereas the presence of vacancies leads to three-dimensional configurations. Additionally, Metropolis Monte Carlo simulations are carried out at high temperature in order to investigate the dependence of nanocluster composition on temperature. A good agreement between the existing experimental data on the ratios (Y+Ti):O, Y:Ti, (Y+Cr):O, and Y:Cr, and the simulation results is found. In some cases it is even possible to draw the conclusion that the respective alloys contained a certain amount of vacancies, and that the clusters analyzed were frozen-in high-temperature configurations. The comparison of experimental data with those obtained by simulations demonstrates that the assumption of nanoclusters consisting of nonstoichiometric oxides which are essentially coherent with the bcc lattice of the Fe-Cr matrix leads to reasonable results.

[*] Corresponding author,

Address: Helmholtz-Zentrum Dresden-Rossendorf, P.O.Box 510119, 01314 Dresden, Germany

Electronic address: M.Posselt@hzdr.de, Phone: +49 351 260 3279, Fax: +49 351 260 3285

PACS numbers: 62.23.St; 61.72.J-; 81.30.Mh; 82.60.-s

## I. INTRODUCTION

Oxide Dispersion Strengthened (ODS) Fe-Cr alloys, or Nanostructured Ferritic Alloys (NFA), consist of a polycrystalline matrix containing bcc crystallites with different populations of tiny oxide-like particles. These materials are produced by mechanical alloying using the powder of a Fe-Cr alloy with small additions of Ti and other metals as well as fine yttria ($Y_2O_3$) powder. Subsequently, the alloy is consolidated by hot isostatic pressing or hot extrusion. Due to the very low solubility of oxygen in the ferritic matrix a standard casting technique cannot be employed to fabricate the ODS alloys. It is assumed that most of the $Y_2O_3$ is dissolved by the mechanical alloying, and during the subsequent thermal processing the oxide-like precipitates are formed.[1] Atom Probe Tomography (APT), Scanning Transmission Electron Microscopy (STEM) in combination with Energy Dispersive X-ray spectroscopy (EDS), Positron Lifetime Spectroscopy (PLS), Small Angle Neutron Scattering (SANS), and other advanced analytical techniques are employed to understand the properties of the small particles in the ODS alloy. Many details of their structure and composition are not yet fully understood. At least two distinct classes of tiny particles were found: (i) stoichiometric oxides ($Y_2Ti_2O_7$/$Y_2TiO_5$), and (ii) nonstoichiometric nanoclusters that contain mainly O, Y, Ti and that are essentially coherent with the bcc-Fe-Cr matrix.[2-4] Different information on the size of these particles can be found in the literature. According to present knowledge the nanoclusters are smaller (1 - 10 nm) than the oxide particles (5 - 30 nm). The former have a much higher density than the latter and are observed both inside the grains and at grain boundaries.[2] Recent investigations found also $Y_2Ti_2O_7$ particles with a size below 5 nm, which was interpreted as a result that is in contrast to the common assumption of nonstoichiometric and coherent nanocluster.[5,6,7] Numerous experiments demonstrated that number and size of the nanoclusters do not change significantly when ODS alloys are exposed to high dose irradiation and/or high temperatures.[4,8-15] Furthermore, it was shown that the fine dispersion of the nanoclusters prevents recrystallization, i.e. the increase of grain size, which usually occurs at elevated temperatures.[16] The extraordinary properties of the nanoclusters are deemed to be the cause of the superior high-temperature creep strength[17-20] and the high radiation resistance of the ODS Fe-Cr alloys.[21] Therefore, these materials are promising candidates for applications as structural materials in extreme environments, i.e. at high temperature and intense particle irradiation, such as in advanced nuclear fission and fusion reactors.[13-15,21]

Besides advanced structural analysis atomic-scale computer simulations can be very helpful to improve the understanding of the nature of the small oxide-like particles in ODS alloys. First-principle methods such as Density Functional Theory (DFT) are highly accurate but practically only applicable to investigate the properties of very small embedded clusters. The computational effort strongly increases with cluster size since larger supercells must be considered and the number of spatial configurations to be investigated for a cluster of a given size and composition grows considerably. DFT calculations yielded valuable fundamental data on the properties of single foreign atoms, point defects, and clusters consisting of few atomic species.[7,21-25] In this work an approach within the framework of a rigid lattice model is employed in order to investigate larger clusters. The interactions between the different atomic species are described by parameters that are obtained by extensive DFT calculations on small clusters. In contrast to previous studies not only parameters for pair interactions but also those for triple interactions are determined. In this manner the accuracy of the approach is to be improved. The fabrication of ODS material by mechanical alloying and subsequent consolidation using hot isostatic pressing or hot extrusion is a very complex process and difficult to model. Under simplified conditions the formation of the nanoparticles embedded in the alloy was studied by Kinetic Monte Carlo simulations.[27-29] The present work is focused on the final state of (sub)nanometer-size clusters that should correspond to a certain thermodynamic equilibrium configuration.[9,30] The aim of the study is the determination of the energetics, structure and composition of the nanoclusters that may contain O, vacancies (v), Y, Ti and/or Cr. The present work is based on the assumption that the clusters are essentially coherent with the bcc-Fe-Cr matrix, which is in agreement with most, but not all, interpretations of the experimental data. In the framework of the rigid lattice model Metropolis Monte Carlo (MMC) simulations are performed to determine the equilibrium configuration of the nanoclusters. Two limiting cases are considered: Cluster configurations at $T=0$ and those at a relatively high temperature. The state at $T=0$ is of general interest to determine the cluster binding energies that can be used as input parameters of coarse-grained methods such as object kinetic Monte Carlo simulations and rate theory which are employed to investigate the nanostructure evolution under irradiation and/or thermal load. The results of the MMC calculations are compared with experimental data. The comparison may contribute to answer the open question whether the nanoclusters are nonstoichiometric oxides and essentially coherent with the bcc-Fe-Cr matrix or not. Furthermore, the comparison is used to find out whether or not the state of the nanoclusters obtained by the different methods of structural analysis corresponds to a frozen-in high-temperature state, and to estimate the role of v in the formation of the clusters.

## II. COMPUTATIONAL METHOD

The Vienna ab-initio simulation package (VASP)[31,32] was used to perform DFT calculations on more than 160 different coherent cluster configurations with an average size of 4 and a maximum size of 12 atomic species. In the following the term "atomic species" is always used in order to denote the foreign atoms O, Y, Ti, and Cr as well as the vacancies (v). All DFT calculations were done for a supercell of 128 atoms with periodic boundary conditions applied. Convergence tests were carried out both for Brillouin zone sampling and sufficiency of plane wave basis sets until the total energy change per atom is less than 0.001 eV. The finally chosen values for Brillouin zone sampling and for plane wave energy cut-off are 3x3x3 k points and 500 eV, respectively. For some larger clusters, convergence tests with respect to the size effect were performed using a 250 atoms supercell. All calculations were done employing the spin polarized formalism, projected augmented wave (PAW) pseudopotentials and the GGA-PBE parameterization of the generalized gradient approximation.[33,34] The lattice parameter of pure ferromagnetic Fe is found to be 2.83 Å. Previous DFT studies showed that in bcc-Fe the single v and single Y, Ti, and Cr atoms occupy lattice sites whereas the single O atom prefers the octahedral interstitial site.[22-24,26,27] Due to these findings, in this work each VASP simulation was started with an embedded cluster composed of O atoms on octahedral sites and v, Y, Ti, and Cr on bcc lattice sites. During a DFT calculation the positions of the ions were relaxed while the supercell volume and shape were held constant. The final result is an atomic configuration that corresponds to the minimum of the total energy of the supercell with the cluster. The ionic relaxation was performed using the conjugate gradient algorithm with a force convergence criterion of 0.001 eV/Å. Previous DFT data suggest a strong binding between O and v leading to a supersaturation of O in ODS alloys, contrary to the low solubility of O in pure bcc-Fe.[22,24] Present DFT results show that the substitutional solute atoms Y and Ti have a strong affinity to the O-v pair resulting in small O-v-Y-Ti clusters which are coherent precursors of larger clusters. The detailed discussion of the more than 160 different configurations investigated by VASP simulations will be the subject of a future publication.

The total binding energy of a cluster consisting of $n$ atomic species is defined by

$$E_{bind} = E(X_1 + X_2 + ... + X_n) + (n-1)E^0 - \sum_{i=1}^{n} E(X_i) \qquad (1)$$

$E(X_1 + X_2 + ... + X_n)$ and $E(X_i)$ denote the total energy of the supercells with the cluster $X_1 + X_2 + ... + X_n$ and the monomer of species $X_i$, respectively, while $E^0$ is the total energy of a supercell containing solely Fe atoms. By definition the value of $E_{bind}$ is negative if attraction between the atomic species dominates.

In order to treat ODS alloys in the framework of the rigid lattice model, usually two sub-lattices are considered, where Fe, v, Y, Ti, and Cr can occupy the lattice sites and O can occupy the octahedral sites of the related bcc-Fe lattice, cf. Refs. 7, 27, and 36. In each sub-lattice exchanges between sites are permitted whereas exchanges between the sub-lattices are prohibited. In the present work a completely identical representation by an underlying simple cubic lattice is employed, where the bcc unit cell is subdivided into eight cubes. The rules for occupation and exchange are identical to those described above. Note that many sites of the simple cubic lattice cannot be occupied. The use of the simple cubic grid allows a more consistent notation of neighbor distances: The minimum distance between O atoms and between O and the other atomic species is the first neighbor distance, whereas the minimum distance between Fe, v, Y, Ti, and Cr is the third neighbor distance in the simple cubic lattice. In the following interactions between atomic species are considered up to the fourth neighbor distance of the simple cubic lattice.

If only pair interactions between the atomic species are taken into account the total binding energy in the supercell is defined by

$$E_{bind} = \sum_i \frac{1}{2} \sum_{jk} \sum_j m(i,j;jk)\, \varepsilon(i,j;jk) \qquad (2)$$

where the parameters $\varepsilon(i,j;jk)$ describe the interaction between species $i$ and $j$, i.e. O, v, Y, Ti, or Cr, at the $jk$ th neighbor distance of the underlying simple cubic lattice, and $m(i,j;jk)$ are geometrical factors. The reference state with $E_{bind} = 0$ corresponds to the case where foreign atoms and v are distributed completely isolated from each other in the bcc-Fe matrix. Due to this definition Fe atoms need not be considered explicitly. The description of the interaction between the atomic species can be improved by the introduction of an additional term that takes into account corrections due the different neighborhoods of the pairwise interacting atomic species.

$$E_{bind} = \sum_i \frac{1}{2} \sum_{jk} \sum_j m(i,j;jk)\, \varepsilon(i,j;jk) + \sum_i \frac{1}{2} \sum_{jl} \sum_j \sum_{kl} \sum_k n(i,j;jl,k;kl)\delta(i,j;jl,k;kl) \qquad (3)$$

The parameters $\delta(i,j;jl,k;kl)$ describe the influence of the neighbor $k$ on the interaction between $i$ and $j$ at the $jl$ th neighbor distance while $k$ is at the $kl$ th neighbor distance from $i$, and $n(i,j;jl,k;kl)$ is a geometrical factor. The pair and triple parameters obey the symmetry relations

$$\varepsilon(i,j;jk) = \varepsilon(j,i;ik) \qquad (4)$$

and

$$\delta(i,j;jl,k;kl) = \delta(i,k;kl,j;jl) \qquad (5)$$

which leads to similar relations for the geometrical factors, due to Eqs. (2) and (3).

The pair parameters $\varepsilon(i,j;jk)$ in Eq. (2) were set equal to the DFT data for the binding energy between two single atomic species in the bcc-Fe matrix. The triple parameters were obtained in the following manner. First, for a certain cluster in bcc-Fe the positions of the atomic species calculated by DFT and mapped to the rigid lattice were used together with Eq. (3) to determine the geometrical factors $m(i,j;jk)$ and $n(i,j;jl,k;kl)$. In this step the parameters $\delta(i,j;jl,k;kl)$ can be chosen arbitrarily. Second, the values of $\delta(i,j;jl,k;kl)$ were fitted to DFT data for $E_{bind}$ using Eq. (3). Table I shows values of $\varepsilon(i,j;jk)$ for all 32 possible pairwise combinations, whereas only 26 values of $\delta(i,j;jl,k;kl)$ are given in Table II. The other triple parameters were set to zero. Of course, the number of nonzero values can be increased by including more DFT results in the fit. In this work a sort of educated guess was employed in order to select the most suitable cluster configurations for the fit. The accuracy of the parameterization by $\varepsilon(i,j;jk)$ and $\delta(i,j;jl,k;kl)$ was estimated as described in the following. The value of $E_{bind}$ determined by Eqs. (2) and (3) was compared with that obtained by DFT und the relative error of this quantity was calculated. Averaging over all configurations considered leads to a mean relative error of 0.38 if only the pair parameterization, Eq. (2), is used. If both pair and triple parameterizations, Eq. (3), are employed the mean relative error decreases to 0.28. In the error estimation only those 125 cluster configurations were included for which DFT calculations yield negative values of $E_{bind}$, i.e. where the attraction between the atomic species dominates. The fact that the use of both pair and triple interaction parameters leads to more precise results demonstrates the influence of the neighborhood on pairwise interacting atomic species if these species belong to a cluster. The present parameterization of the interaction between the atomic species is much more accurate than that used in previous studies which also employed the rigid lattice model to treat ODS alloys.[27,28,36] In the case of the pair parameterization given in Ref. 27, that does not consider clusters containing Ti, the error estimation described above yields a mean relative error of 1.03. There are several reasons for the still considerable mean relative error of the present parameterization. In the framework of the rigid lattice model the relaxation of atomic positions due to the internal strain, e.g. because of size mismatch between the different atomic species, cannot be considered explicitly. Instead the relaxation is taken into account implicitly, since the interaction parameters $\varepsilon(i,j;jk)$ and $\delta(i,j;jl,k;kl)$ were fitted to DFT reference data that bear in mind the relaxation of the positions of atoms. DFT calculations yield a strong relaxation of the O-v and Y-v pairs which were initially placed at first and third neighbor positions of the simple cubic lattice, cf. also results shown Refs. 22, 24 and 26. While the present parameterization describes the relaxation due to the interaction between two single atomic species correctly, see above, the treatment of the interactions in clusters with more than two atomic species is less precise. Another point that can be taken into account only approximately by the fit to a certain number of DFT reference data is the partial ionic character of the bonds in the oxidic nanoclusters (cf. Ref. 23), including the effects of charge redistribution and compensation.

The general trends concerning the strength of attractive interactions between the atomic species are illustrated in Fig.1. This complex scheme is broadly consistent with results obtained by other authors.[22-24,26-28] Fig. 1 reveals the strong difference between the presence and the absence of v. Without any v, the most attractive interaction is that between O and Y, followed by that between O and Ti, in both cases at the second neighbor distance of the underlying simple cubic lattice. A peculiarity is the dominating attraction between two O atoms at the fourth neighbor distance if there is no occupied bcc lattice site between them. If v are present the most important attractive interaction is that between O and v, at the first neighbor distance, followed by that between v and Y at the third neighbor distance while the attractions between v and Ti and between two v are weaker. Cr shows only a rather weak attraction to v. On the whole, its influence on the formation of the nanoclusters should be small.

In the MMC simulations the following, well-known algorithm is employed. First, a certain number of atomic species is randomly distributed, where Fe atoms are replaced by v, Y, Ti, or Cr while O occupies the octahedral sites of the related bcc lattice. Then, v, Y, Ti, and Cr atoms are randomly exchanged with unequal species (including Fe) on bcc lattice sites and the distribution of O atoms on the octahedral sites is also randomly changed. After the exchange/change the energy of the old and the new configurations is compared. If the energy of the new state $n$ is lower than that of the old state $o$, the new configuration is accepted. In the case of an increase of energy the new configuration is accepted with a probability proportional to $\exp\left(-\dfrac{(E_n - E_o)}{k_B T}\right)$, where $T$ and $k_B$ are the current temperature and the Boltzmann constant, respectively. At a given temperature all the possible exchanges or changes are performed $mmcs_{max}$ times, until the supercell reaches a steady state, i.e. its free energy evolves towards its minimum. In the present work the quantities $E_o$ and $E_n$ correspond to the respective total binding energies of the system. In order to determine the most stable cluster configuration at $T=0$ and the corresponding total binding energy the method of simulated annealing (SA) is used. It consists of numerous MMC simulation steps at different temperatures. Starting with $T_{max}$ the temperature is decreased in steps of $\Delta T$ until $T=0$ is reached. The reliability of the MMC and SA algorithms must be checked by performing several simulations with different seed numbers of the random number generator and with various values of the parameters $T_{max}$, $mmcs_{max}$, and $\Delta T$. Depending on the cluster size ten to thousand independent SA calculations were performed and the configuration with the highest absolute value of the total binding energy was selected. In all SA and MMC calculations a single compact cluster was found. The size of the supercell with the rigid lattice was generally chosen in such a manner that the ratio between foreign atomic species (O, v, Y, Ti, Cr) and Fe is less than about 15:100. The SA and MMC simulations contain an automatic cluster analysis including the determination of the cluster stoichiometry.

### III. RESULTS AND DISCUSSION

The calculation methods described in the last section were applied to investigate nanocluster formation in three typical ODS alloys: (i) a model alloy without Ti and a O:Y ratio of about 18:12 (cf. Ref. 35), (ii) a YWT alloy with the atomic ratio O:Y:Ti of 18:12:46 (cf. Refs. 1, 8, 10, 30, 37, and 38) and (iii) a MA957 alloy with the atomic ratio O:Y:Ti of 18:12:100 (cf. Refs. 4, 35, and 37). YWT alloys were produced in different laboratories, whereby the acronym indicates alloying additions of $Y_2O_3$, W, and Ti, and numbers in front, e.g. in 14YWT and 12YWT, give the weight percent of Cr concentration. MA957 is a commercial product with a similar composition as 14YWT, but it contains Mo instead of W. Previous theoretical and experimental investigations showed that v introduced during the fabrication of the ODS material by mechanical alloying enables the incorporation of a relatively high amount of O and should have an important influence on the formation and properties of the nanoclusters.[2,9,22,30,39] In order to investigate this effect supercells with different concentrations of v were studied: (i) no v in the supercell, and O:v ratios of about (ii) 4:1, (iii) 2:1, and (iv) 1:1. As already mentioned in the last paragraph, the effect of Cr on formation and properties of the nanoclusters should be relatively small, although the Cr content in the ODS alloy is high. Therefore, the role of Cr is studied separately. The influence of other minor alloying elements (W, Mo, etc.) and additives on the formation and properties of the nanoclusters is deemed to be negligible.

The following description starts with the results on structure and energetics of the most stable clusters at $T = 0$ obtained by SA calculations, and continues with the discussion of data on the composition of the nanoclusters at $T = 0$ and at a high temperature determined by SA and MMC simulations, respectively.

Fig. 2 depicts the total binding per O atom in the cluster, $E_{b/O}$, for nanoclusters containing 3 to 21 oxygen atoms, together with characteristic cluster configurations. The supercell does not contain any v. Fig. 2 shows that with increasing cluster size the gain of total binding energy decreases continuously. This trend is smoother and somewhat more pronounced if the atomic interaction is described by both pair and triple parameters. Results of SA calculations using pair parameters are shown for comparison, but only for clusters containing 3 to 10 O atoms. The less accurate pair parameterization yields higher absolute values of $E_{b/O}$. The relative gain of total binding energy is generally smaller if Ti is present. In all cases considered the clusters show a planar structure. Obviously, this is due to the peculiarity that the dominating attraction between two O atoms is that at the fourth neighbor distance in the underlying simple cubic lattice. It should be mentioned that the small O-Y-Ti clusters investigated by DFT calculation show also such a planar structure. This is in complete agreement with recent DFT results of Claisse *et al.*[26] They also found that the planar structure of a cluster containing three O, two Y and one Ti is more stable than the corresponding three-dimensional structure proposed in Ref. 23. The comparison of the data for the two different Ti concentrations shows that more Ti in the supercell does not lead to more Ti in the cluster. In general the same or very similar cluster configurations are observed for the atomic ratios O:Y:Ti ≈ 18:12:46 and O:Y:Ti ≈ 18:12:100. The excess of Ti atoms is distributed in the bcc-Fe matrix, completely isolated from each other. O is always inside the cluster. The stoichiometry of the clusters will be discussed in detail later.

The effect of v on energetics and structure of the nanoclusters is illustrated in Fig. 3, for the case that the number of v in the supercell equals the number of O atoms. The absolute value of the total binding energy per O atom is generally higher than in the case without v. Again, the gain of total binding energy decreases with cluster size and the relative gain is lower if Ti is present. Calculations using the more precise pair and triple parameterization lead to lower absolute values of $E_{b/O}$ than those using only pair parameters. The more accurate calculations lead to a stronger and more continuous decrease of the gain of binding energy vs. the number of O atoms, $n_O$, in the cluster. One might even assume saturation at large clusters sizes. The trends illustrated both in Fig. 2 and Fig. 3 are fully consistent with the observation of a small average size and a high dispersion of the nanoclusters in ODS alloys as well as with the role of Ti as an inhibitor of cluster growth.[1,21,40] The strong increase of the absolute value of the binding energy if v are added to the cluster is consistent with the observation of the high radiation tolerance of ODS Fe-Cr alloys. The presence of v leads to a three-dimensional structure of the clusters, cf. Fig. 3, which is in agreement with the results of DFT calculations for small clusters. O and v are always part of the cluster and are arranged frequently in parallel O-v chains. Again, the same or a very similar cluster configuration is observed for the atomic ratios O:Y:Ti ≈ 18:12:46 and O:Y:Ti ≈ 18:12:100. That means that only a certain number of Ti atoms can be incorporated into a cluster of a given number of O atoms. A similar limit should also exist for Y atoms, cf. Fig. 3. The excess Ti and Y atoms are thinly dispersed in the matrix. In the case of a cluster containing 15 O atoms, Fig. 4 illustrates that the addition of v leads to a transition from a completely planar structure to a fully three-dimensional configuration. It can be assumed that already one v leads to a three-dimensional substructure. For the supercells with O:v ratios of about 4:1 and 2:1 the absolute values of the corresponding total binding energy per O atom (not shown) lie between those for supercells without v and those for supercells where the number of v equals the number of O atoms. Note that the results presented in Fig. 4 and in the following figures were obtained by calculations using both pair and triple parameters to describe the interaction between the atomic species.

The influence of Cr was studied for the three different alloy compositions and for the cases without and with v. In the calculations the number of Cr atoms in the supercell was chosen sufficiently high in order to provide enough Cr for the binding to or the incorporation into a cluster. However, it was found that Cr is only part of the clusters in the alloy with the atomic ratio O:Y ≈ 18:12 if v are present. Fig. 5 depicts the results of the detailed study for different numbers of v in the supercell. A few Cr atoms can reside at the cluster periphery but the value of $E_{b/O}$ is only slightly influenced by the presence of Cr. This is due to the weak attractive interaction of Cr with the other atomic species. Cr is not part of a cluster if Ti is present, independently of the amount of v. The simulation results agree very well with experimental data. It was found[35] that in the model alloy with the atomic ratio O:Y ≈ 18:12 the nm-size clusters have a Cr content comparable to that of O. On the other hand, the Cr content obtained for the nanoclusters in alloys containing Ti is much lower.[1,8,10,11,35,37,41] Moreover,

Fig. 5 illustrates that in the alloy with the atomic ratio O:Y $\approx$ 18:12 Cr atoms form the shell of the nanoclusters and the core consists of O, v, and Y, which is in agreement with APT data of Marquis *et al.*[14,35]

In the case of relatively large clusters containing about 100 atomic species up to 1000 independent SA calculations, starting with different initial random numbers, had to be performed in order to get reliable results. This is because of the existence of a large number of configurations with very similar binding energies, which may be also related to the lack of a significant driving force for cluster growth as discussed above. Furthermore, it should be noticed that although presentations of cluster configurations show the atomic species on the sites of a rigid lattice, in all calculations off-lattice relaxations were taken into account implicitly, as discussed in section II. In particular this concerns the relaxation in the neighborhood of v. Therefore, in the presence of v the nanocluster structure is really not completely coherent with the bcc matrix.

The composition of the nanoclusters at $T = 0$ was obtained from the SA results whereas the stoichiometry data for 1687 K were determined by separate MMC simulations. The two extreme temperatures were chosen in order to investigate a possible dependence of composition on temperature. On the one hand, the value of 1687 K is simply chosen to obtain characteristic high-temperature results. On the other hand, 1687 K is presumably the highest annealing temperature of ODS alloys reported in literature.[9]

Fig. 6 shows the stoichiometry of the nanoclusters in a YWT-type alloy. If the alloy contains v the ratio (Y+Ti):O is higher than in the case without v, which can be explained by the higher absolute value of the total binding energy, cf. Fig. 3. At the high temperature the ratio (Y+Ti):O in the cluster is lower than at $T = 0$ and the size dependence is weaker, both in the case with and without v in the supercell (Fig. 6a). The temperature dependence of the ratio (Y+Ti):O is due to the fact that at high temperature the number of free Ti atoms in the matrix is higher than at $T=0$, due to the only modest binding of Ti to O and v (cf. Fig. 1). The decrease of (Y+Ti):O with cluster size can be explained by the decreasing surface-to-volume ratio since O atoms are solely located in the interior of the cluster. The Y:Ti ratio presented in Fig. 6b increases/decreases with temperature in cases with/without v in the supercell. If v are present the temperature dependence is again caused by the only modest binding of Ti to O and v. If the system does not contain v, the cluster exhibits a planar structure, as discussed above. The analysis of results of SA and MMC simulations shows that in this case the ratio of free Y to free Ti atoms at the high temperature is higher than at $T=0$. For supercells with O:v ratios of about 4:1 and 2:1 the results (not shown) range between the case without v and those with a O:v ratio of 1:1. Fig. 6 demonstrates that the results of SA and MMC simulations lie within the same range as the APT data obtained by different authors.[1,8,10,11,30,37] Two APT results[1,8] might be even attributed to the fact that in these samples the clusters contain a relatively small amount of v.

The stoichiometry of the nanoclusters in a MA957-type alloy is illustrated in Fig. 7. The atomic ratios (Y+Ti):O and Y:Ti show trends similar to those displayed in Fig. 6. This is consistent with the data on the total binding energy per O atom shown in Figs. 3b and 3c. The small differences between Figs. 6 and 7 might result from the fact that in the latter case the clusters contain slightly more Ti atoms. The simulations results of Fig. 7 agree well with the APT data of Refs. 35 and 37 . The comparison indicates that the clusters analyzed by APT do not contain too much v. Moreover, it can be assumed that the APT data of Refs. 1, 35, and 37 depicted in Figs. 6 and 7 result from cluster configurations that were frozen in at the end of the hot consolidation process. The arrow on the right-hand side of Fig. 7b illustrates the range of data for clusters with a size of less than 10 nm, obtained by Field Emission Gun-Transmission Electron Microscopy (FE-TEM) with an Energy Dispersive X-ray Spectrometry system (EDS).[4] These experimental data lie within the same range as the results predicted by simulations.

The maximum cluster size considered in the simulations is about 100 atomic species which corresponds to diameters of about 1 and 2 nm if purely three-dimensional and purely two-dimensional cluster are considered, respectively. In general these diameters are by a factor 2-4 lower than the mean size of the nanoclusters investigated experimentally. However, the results shown in Figs. 6 and 7 reveal that the size dependence of the ratios (Y+Ti):O and Y:Ti becomes weaker with increasing cluster size. Thus, the comparison of present simulation results with the APT data is justified.

Fig. 8 presents results of SA and MMC simulations for the model alloy that does not contain Ti. The size and temperature dependence of the ratios (Y+Cr):O and Y:Cr is similar to that of the ratios (Y+Ti):O and Y:Ti depicted in Figs. 6 and 7. Note that at $T=0$ Cr is only part of a cluster if v are present. However, at the high temperature clusters can contain some Cr atoms in thermal equilibrium, even if no v are present. The APT data of Marquis *et al.*[35] are closest to the simulation results for $T=0$, for a supercell that contains as much v as O atoms. This may be an indication that in alloys without Ti the nanoclusters contain more v than in those with Ti, and that there is some cluster evolution during cooling after hot consolidation.

## IV. SUMMARY AND CONCLUSIONS

In the framework of a rigid lattice model SA and MMC calculations were performed to determine the energetics, structure and composition of nanoclusters in three different ODS Fe-Cr alloys. The interaction between the atomic species was treated by parameters obtained from comprehensive DFT calculations on very small clusters. It was shown that a novel description using both pair and triple parameters is more precise than the commonly applied pair parameterization. The triple parameters imply corrections due to the different neighborhoods of the pairwise interacting atomic species. The complex scheme of the dominating attractive interactions is consistent with results of previous theoretical investigations. It reveals a strong difference between the presence and absence of v in the alloy. In the former and the latter case the binding between O and v and between O and Y is the strongest, respectively. A peculiarity is the dominating attraction between O atoms at the fourth neighbor distance of the underlying simple cubic lattice.

The SA calculations provided various data on the properties of the nanoclusters at $T=0$. The results show that with increasing cluster size the gain of total binding energy per O atom decreases continuously. The relative gain is lower if Ti is present in the alloy. The same trends are observed if the nanoclusters contain v, but in this case the absolute value of the binding energy is considerably higher. These results are fully consistent with the experimental observation of a small average size and a high dispersion of the nanoclusters in ODS alloys, and with the finding that the presence of Ti leads to a reduction of nanocluster size. The considerable increase of the absolute value of the binding energy if v are added to the cluster can be related to the high radiation tolerance of ODS Fe-Cr alloys. The calculations showed that clusters with and without v have a three-dimensional and a planar structure, respectively. For a given number of O atoms, there exists a limit for the incorporation of Ti and Y atoms into the cluster, whereas O and v are always part of the cluster. The attractive interaction of Cr with the other atomic species is relatively weak. Thus, Cr is not part of the nanoclusters, except for alloys without Ti but with v. In the latter case the clusters consist of a core containing O, v, and Y and a Cr shell, which is in agreement with experimental findings.

The stoichiometry of the nanoclusters at $T=0$ and at 1687 K was obtained from the results of SA calculations and from separate MMC simulations, respectively. The two extreme temperatures were chosen in order to study the dependence of nanocluster composition on temperature. If the alloy contains v, the atomic ratios (Y+Ti):O and (Y+Cr):O in the cluster are higher than in the case without v, since, due to the higher absolute value of the total binding energy per O atom, more Ti and Cr atoms bound. For the same reason Y:Ti and Y:Cr decrease with increasing v content. The ratios (Y+Ti):O and (Y+Cr):O decrease with cluster size, whereby the size dependence becomes generally weaker. The overall comparison of the results for $T = 0$ and 1687 K reveals the high thermal stability of the nanoclusters which is in agreement with experimental findings. At 1687 K the ratios (Y+Ti):O and (Y+Cr):O are lower than at $T = 0$ and the size dependence is weaker. This is due to the fact that at high temperature more Ti and Cr atoms are dissolved in the matrix. The same reason leads to the temperature dependence of the ratios Y:Ti and Y:Cr if the alloy contains v. On the other hand, Y:Ti decreases with temperature if no v are in the material. For the three different alloys considered the simulation results lie within the range of numerous experimental data on the nanocluster composition. In several cases it is even possible to draw the conclusion that the respective alloys contained more or less v, or that the experimental data result from cluster configurations that were frozen in at the end of the hot consolidation process.

The comparison of the data determined by SA and MMC simulations with qualitative and quantitative experimental results demonstrates that the assumption of nanoclusters consisting of nonstoichiometric oxides which are essentially coherent with the bcc lattice of the Fe-Cr matrix leads to reasonable results.

In conclusion, it should be emphasized that the present multiscale modeling approach yields, for the first time, a comprehensive understanding of existing experimental data on structure and stoichiometry of the nanoclusters in ODS alloys.

# References

[1] M. J. Alinger, G. R. Odette, and D. T. Hoelzer, Acta Mater. **57**, 392 (2009).

[2] M. K. Miller, C. M. Parish, and Q. Li, Mater. Sci. Technolog. **29**, 1174 (2013).

[3] M. C. Brandes, L. Kovarik, M. K. Miller, and M. J. Mills, J. Mater. Sci. **47**, 3913 (2012).

[4] H. Sakasegawa, F. Legendre, L. Boulanger, M. Brocq, L. Chaffron, T. Cozzika, J. Malaplate, J. Henry, and Y. de Carlan, J. Nucl. Mater. **417**, 229 (2011).

[5] Y. Wu, E. M. Haney, N. J. Cunningham, and G. R. Odette, Acta Mater. **60**, 3456 (2012).

[6] S. Y. Zhong, J. Ribis, V. Klosek, Y. de Carlan, N. Lochet, V. Ji, and M. H. Mathon, J. Nucl. Mater. **428**, 154 (2012).

[7] L. Barnard, G. R. Odette, I. Szlufarska, and D. Morgan, Acta Mater. **60**, 935 (2012).

[8] M. K. Miller, E. A. Kenik, K. F. Russell, L. Heatherly, D. T. Hoelzer, and P. J. Maziasz, Mater. Sci. Eng. A **353**, 140 (2003).

[9] X. L. Wang, C. T. Liu, U. Keiderling, A. D. Stoica, L. Yang, M. K. Miller, C. L. Fu, D. Ma, and K. An, J. Alloys Compd. **529**, 96 (2012).

[10] M. K. Miller, K. F. Russell, and D. T. Hoelzer, J. Nucl. Mater. **351**, 261 (2006).

[11] P. Pareige, M. K. Miller, R. E. Stoller, D. T. Hoelzer, E. Cadel, and B. Radiguet, J. Nucl. Mater. **360**, 136 (2007).

[12] S. Yamashita, N. Akasaka, and S. Ohnuki, J. Nucl. Mater. **329-333**, 377 (2004).

[13] G. R. Odette, M. J. Alinger, and B. D. Wirth, Annu. Rev. Mater. Res. **38**, 471 (2008).

[14] E. A. Marquis, J. M. Hyde, D. W. Saxey, S. Lozano-Perez, V. de Castro, D. Hudson, C. A. Williams, S. Humphry-Baker, and G. D. W. Smith, Mater. Today **12**, 30 (2009).

[15] G. R. Odette and D. T. Hoelzer, JOM **62**, 84 (2010).

[16] H. R. Z. Sandim, R. A. Renzetti, A. F. Padilha, D. Raabe, M. Klimenkov, R. Lindau, and A. Möslang, Mater. Sci. Eng. A **527**, 3602 (2010).

[17] R. L. Klueh, P. J. Maziasz, I. S. Kim, L. Heatherly, D. T. Hoelzer, N. Hashimoto, E. A.

Kenik, and K. Miyahara, J. Nucl. Mater. **307-311**, 773 (2002).

[18] R. L. Klueh, J. P. Shingledecker, R. W. Swindeman, and D. T. Hoelzer, J. Nucl. Mater. **341**,103 (2005).

[19] J. H. Schneibel, C. T. Liu, M. K. Miller, M. J. Mills, P. Sarosi, M. Heilmaier, and D. Sturm, Scr. Mater. **61**, 793 (2009).

[20] M. B. Toloczko, F. A. Garner, and S. A. Maloy, J. Nucl. Mater. **428**, 170 (2012).

[21] S. Ukai, S. Mizuta, T. Yoshitake, T. Okuda, M. Fujiwara, S. Hagi, and T. Kobayashi, J. Nucl. Mater. **283-287**, 702 (2000).

[22] C. L. Fu, M. Krčmar, G. S. Painter, and X.-Q. Chen, Phys. Rev. Lett. **99**, 225502 (2007).

[23] Y. Jiang, J. R. Smith, and G. R. Odette, Phys. Rev. B **79**, 064103 (2009).

[24] D. Murali, B. K. Panigrahi, M. C. Valsakumar, S. Chandra, C. S. Sundar, and B. Raj, J. Nucl. Mater. **403**,113 (2010).

[25] H. Zhao, C. L. Fu, M. Krčmar, and M. K. Miller, Phys. Rev. B **84**, 144115 (2011).

[26] A. Claisse and P. Olsson, Nucl. Instrum. Meth. Phys. Res. B **303**, 18 (2013).

[27] C. Hin, B. D. Wirth, and J. B. Neaton, Phys. Rev. B **80**, 134118 (2009).

[28] P. Jegadeesan, D. Murali, B. K. Panigrahi, M. C. Valsakumar, and C. S. Sundar, Int. J. Nanosci. **10**, 973 (2011).

[29] C. Hin and B. D. Wirth, Mater. Sci. Eng. A **528**, 2056 (2011).

[30] M. K. Miller, C. L. Fu, M. Krčmar, D. T. Hoelzer, and C. T. Liu, Front. Mater. Sci. China **3**, 9 (2009).

[31] G. Kresse and J. Furthmüller, Phys. Rev. B **54**, 11169 (1996).

[32] G. Kresse and D. Joubert, Phys. Rev. B **59**, 1758 (1999).

[33] P. E. Blöchl, C. J. Först, and J. Schimpl, B. Mater. Sci. **26**, 33 (2003).

[34] J. P. Perdew, K. Burke, and M. Ernzerhof, Phys. Rev. Lett. **77**, 3865 (1996).

[35] E. A. Marquis, Appl. Phys. Lett. **93**, 181904 (2008).

[36] M. J. Alinger, B. D. Wirth, H.-J. Lee, and G. R. Odette, J. Nucl. Mater. **367-370**, 153 (2007).

[37] M. K. Miller, D. T. Hoelzer, E. A. Kenik, and K. F. Russell, Intermetallics **13**, 387 (2005).

[38] A. Hirata, T. Fujita, C. T. Liu, and M. W. Chen, Acta Mater. **60**, 5686 (2012).

[39] J. Xu, C. T. Liu, M. K. Miller, and H. Chen, Phys. Rev. B **79**, 020204(R) (2009).

[40] S. Ukai and M. Fujiwara, J. Nucl. Mater. **307-311**, 749 (2002).

[41] M. K. Miller, D. T. Hoelzer, E. A. Kenik, and K. F. Russell, J. Nucl. Mater. **329-333**, 338 (2004).

TABLE I. The 32 pair interaction parameters $\varepsilon(i,j;jk)$ (in eV) defined on the simple cubic lattice, cf. Eq. (2). The parameter $\varepsilon(O,O;4)$ is equal to -0.10 eV if there is no occupied bcc lattice site between the two O atoms. Otherwise it is set to zero. Note the rules for the occupation of lattice sites which are explained in the text.

| I | j | jk | | | |
|---|---|---|---|---|---|
| | | 1 | 2 | 3 | 4 |
| O | O | 0.66 | 0.44 | -0.05 | -0.10/0.0 |
| O | v | -1.65 | -0.75 | | |
| O | Y | 0.35 | -1.01 | | |
| O | Ti | -0.26 | -0.55 | | |
| O | Cr | -0.25 | 0.02 | | |
| V | v | | | 0.15 | -0.25 |
| V | Y | | | -1.45 | -0.26 |
| V | Ti | | | -0.26 | 0.16 |
| V | Cr | | | -0.05 | -0.02 |
| Y | Y | | | 0.19 | 0.01 |
| Y | Ti | | | 0.15 | 0.01 |
| Y | Cr | | | 0.16 | 0.15 |
| Ti | Ti | | | 0.23 | 0.13 |
| Ti | Cr | | | 0.15 | 0.11 |
| Cr | Cr | | | 0.21 | 0.13 |

TABLE II. The 26 nonzero triple interaction parameters $\delta(i,j;jl,k;kl)$ used in the rigid lattice model, cf. Eq. (3).

| $i$ | $j$ | $jl$ | $k$ | $kl$ | $\delta(i,j;jl,k;kl)$ (eV) |
|---|---|---|---|---|---|
| O | O | 1 | O | 1 | 3.1 |
| O | O | 1 | O | 2 | -0.03 |
| O | O | 1 | v | 1 | 3.1 |
| O | O | 1 | v | 2 | 3.1 |
| O | O | 1 | Y | 1 | 3.1 |
| O | O | 1 | Y | 2 | 3.1 |
| O | O | 1 | Ti | 1 | 3.1 |
| O | O | 1 | Ti | 2 | 3.1 |
| O | O | 2 | O | 2 | -0.11 |
| O | O | 2 | v | 1 | 0.33 |
| O | O | 2 | v | 2 | 0.19 |
| O | O | 2 | Y | 2 | 0.31 |
| O | O | 2 | Ti | 1 | 1 |
| O | O | 2 | Ti | 2 | -0.01 |
| O | v | 1 | v | 2 | 0.28 |
| O | Y | 2 | Y | 2 | 0.6 |
| O | Y | 2 | Ti | 1 | 0.08 |
| V | O | 1 | O | 1 | 0.07 |
| V | O | 1 | O | 2 | 0.18 |
| V | O | 2 | O | 2 | -0.05 |
| V | Y | 3 | Y | 3 | 0.76 |
| Y | O | 1 | O | 1 | 3.03 |

| Y | O | 1 | O | 2 | 0.39 |
| Y | O | 2 | O | 2 | -0.19 |
| Ti | O | 1 | O | 1 | 2.02 |
| Ti | O | 1 | O | 2 | -0.03 |

**Figure Captions**

| | |
|---|---|
| Fig. 1 | (Color online) The dominating attractive pair interactions between O, v, Y, Ti, and Cr in a Fe matrix. The arrow thickness indicates the intensity of the interaction at the corresponding distance (1$^{st}$ neighbor – 1nn, 2$^{nd}$ neighbor – 2nn, etc.) in the underlying simple cubic lattice. If the atomic species are not connected by arrows, their interaction is repulsive. The occupation rules for the lattice are explained in the text. |
| Fig. 2 | (Color online) Total binding energy per O atom in a nanocluster, $E_{b/O}$, for clusters containing 3 to 21 oxygen atoms, together with characteristic cluster configurations. The results shown by the red squares were obtained by calculations using both pair and triple parameters, whereas the data depicted by blue circles were determined using only the pair parameters. Thin lines are only drawn to guide the eye. Three atomic ratios typical for ODS alloys were considered: O:Y=18:12 (a), O:Y=18:12:46 (b), O:Y=18:12:100 (c). The cluster configuration with eight O atoms shown in the figure was determined using both pair and triple parameters. Atoms of the bcc-Fe matrix are not shown. In agreement with Fig. 1, O, Y, and Ti atoms are depicted by red, green and yellow spheres, respectively. The formulae below and within the pictures characterize the content of foreign atoms in the supercell and the cluster, respectively. |
| Fig. 3 | (Color online) Total binding energy per O atom in a nanocluster, for alloys containing just as much v as O atoms, together with characteristic cluster configurations. The blue spheres depict v. Bonds are only shown between O atoms as well as between O atoms and other atomic species, in both cases at the 1$^{st}$ neighbor distance; and between the other species at the 3$^{rd}$ neighbor distance of the underlying simple cubic lattice. The quantities shown and the general style of the |
| | presentation are explained in the caption of Fig, 2. |
| Fig. 4 | (Color online) Effect of the v content in the alloy on the spatial configuration of a nanocluster containing 15 O atoms. The presentation style is explained in Fig. 2. |
| Fig. 5 | (Color online) The influence of Cr on the total binding energy per O atom in a nanocluster, $E_{b/O}$, in an alloy with the atomic ratio O:Y=18:12 and different concentrations of v (red and blue symbols). For comparison, the values of $E_{b/O}$ for an alloy without Cr are depicted (black lines). Three examples of spatial nanocluster configurations are shown. Cr atoms are depicted by small black spheres. |
| Fig. 6 | (Color online) Stoichiometry of nanoclusters in the alloy with the atomic ratio O:Y=18:12:46, for materials without v (red symbols and lines), and with the same number of v as O atoms (blue symbols and lines). The symbols depict results of SA calculations whereas the thick lines were obtained from MMC simulations at a temperature of 1687 K. Note that all data were determined for integer numbers of O atoms, the lines in between are only drawn to guide the eye. On the right-hand side of the diagrams experimental APT data from literature are given (orange square: Ref. 11, magenta circle: Ref. 1, green triangle: Ref. 37, gray diamond: Ref. 10, black star: Ref. 30, open diamond with cross Ref. 8). |
| Fig. 7 | (Color online) Composition of nanoclusters in an alloy with the atomic ratio O:Y=18:12:100, for cases without v, and with the same number of v as O atoms. The quantities depicted and the style of the presentation are explained in the caption of Fig. 6. The symbols on the right-hand sides show APT data (magenta square: Ref. 35, green triangle: Ref. 37) while the arrow illustrates the range of data obtained by FE-TEM using EDS.[4] |

| Fig. 8 | (Color online) Stoichiometry of nanoclusters containing Cr, in an alloy with the atomic ratio O:Y=18:12. The magenta squares on the right-hand sides depict the APT data of Marquis *et al.*[35] |
|---|---|

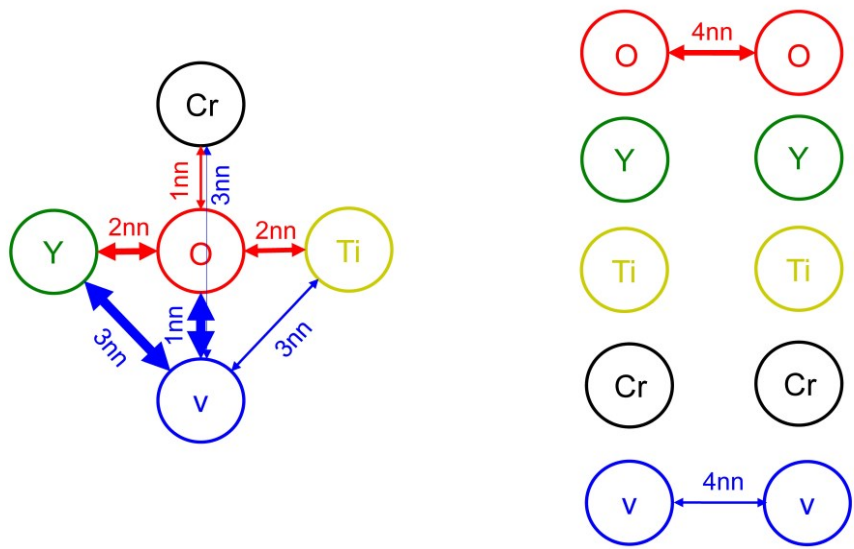

Fig. 1

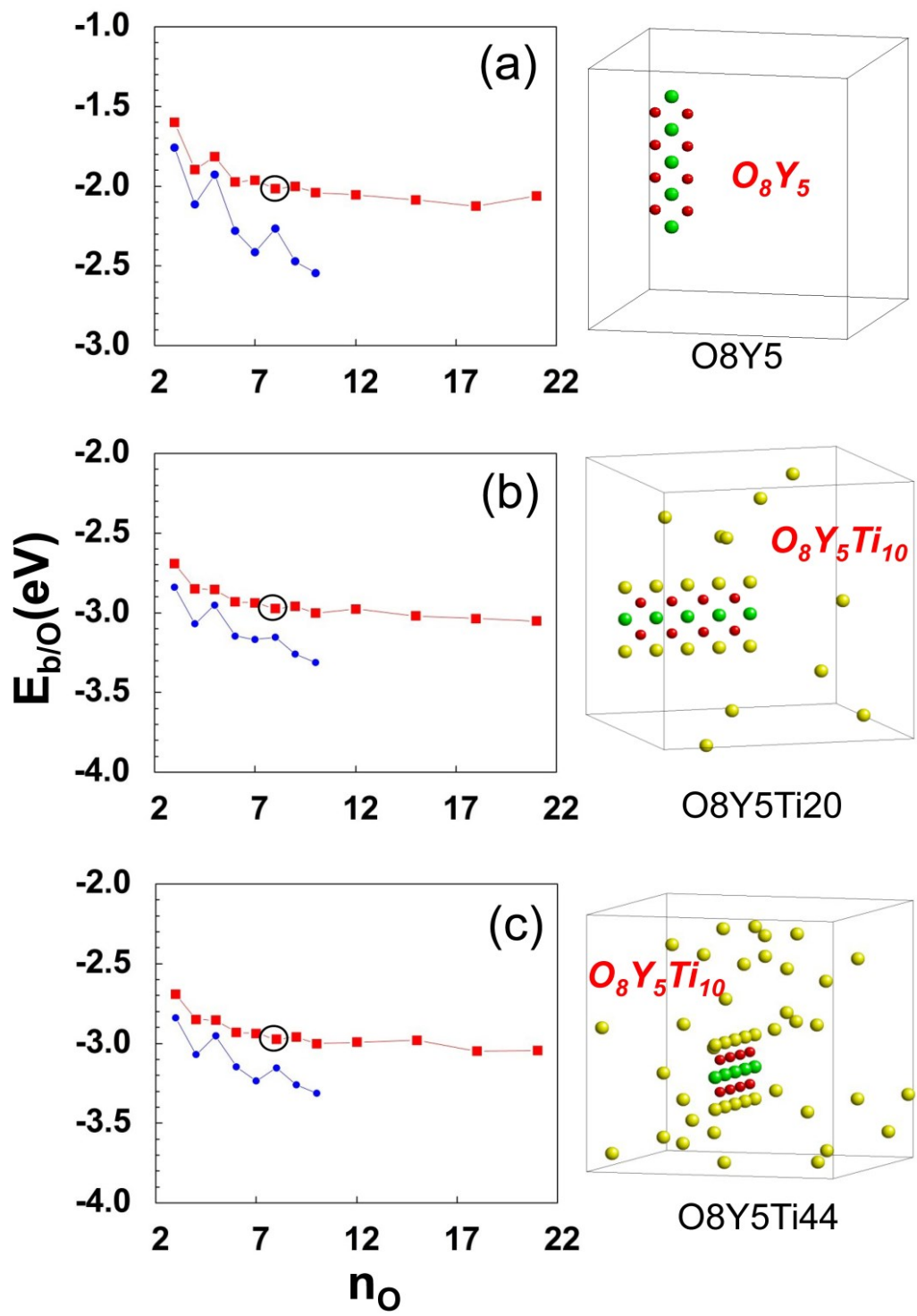

Fig. 2

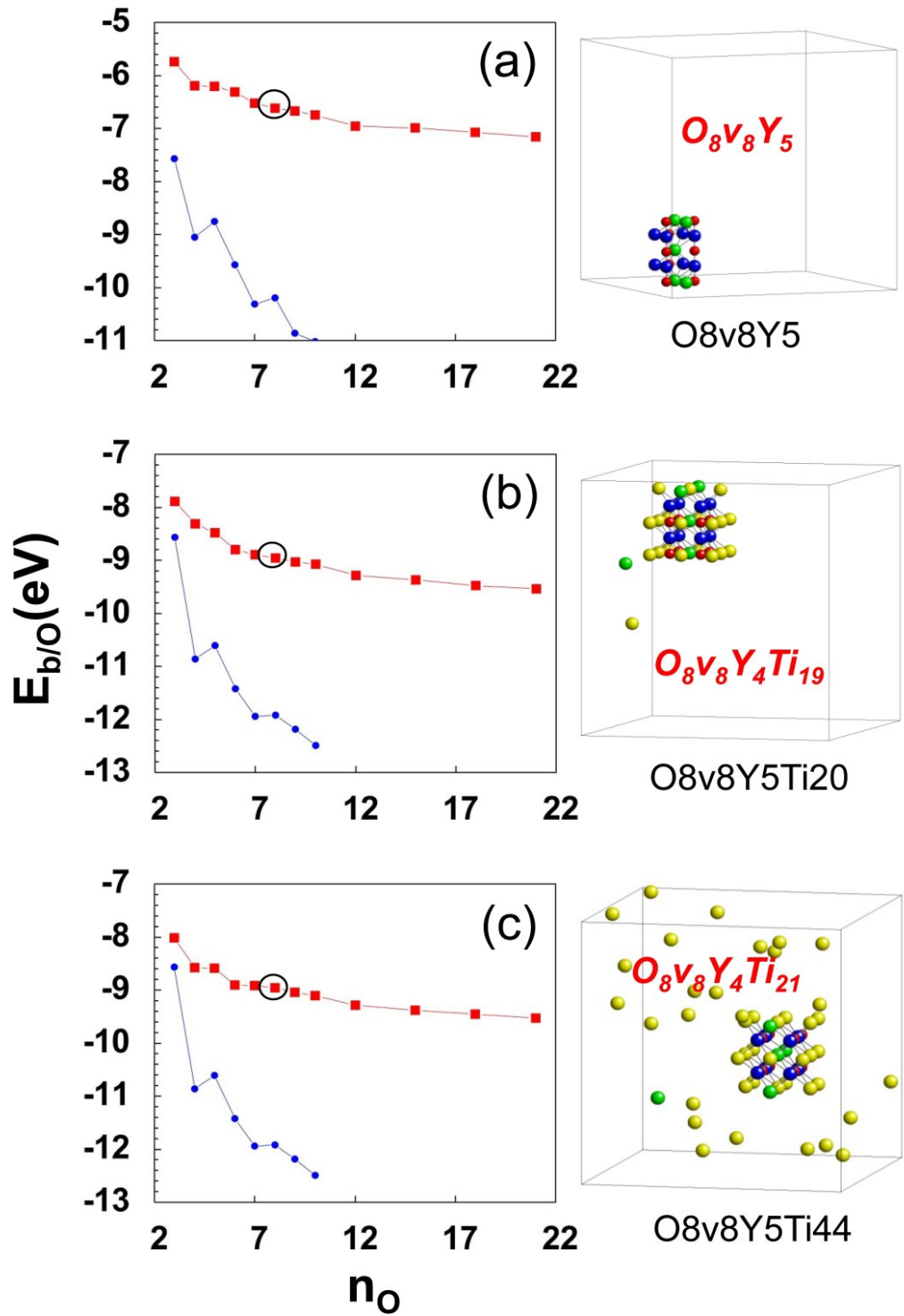

Fig. 3

## $O_{15}Y_{10}Ti_{14}$

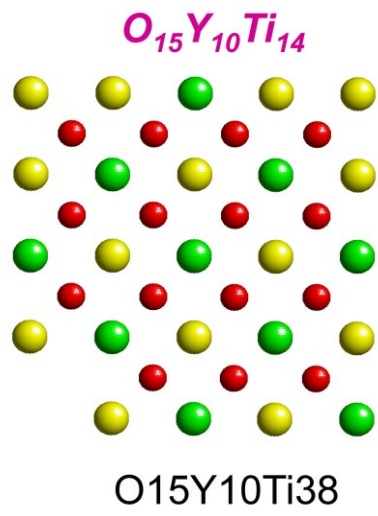

O15Y10Ti38

## $O_{15}v_4Y_9Ti_{21}$

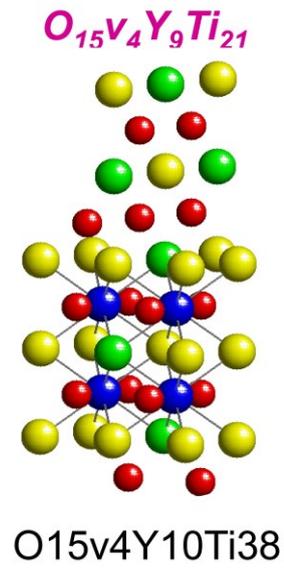

O15v4Y10Ti38

## $O_{15}v_8Y_8Ti_{26}$

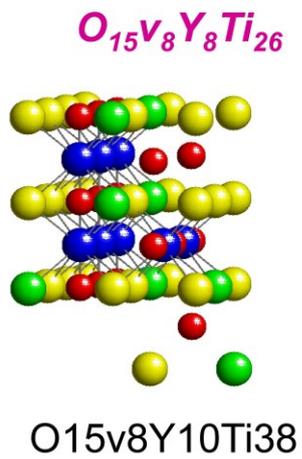

O15v8Y10Ti38

## $O_{15}v_{15}Y_7Ti_{33}$

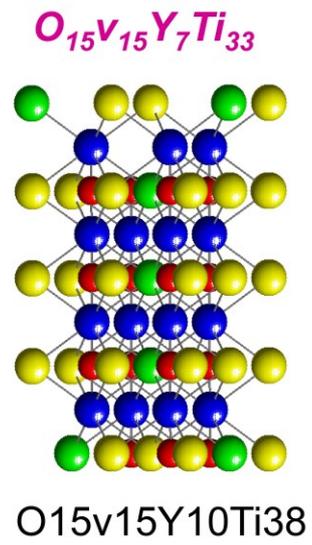

O15v15Y10Ti38

Fig. 4

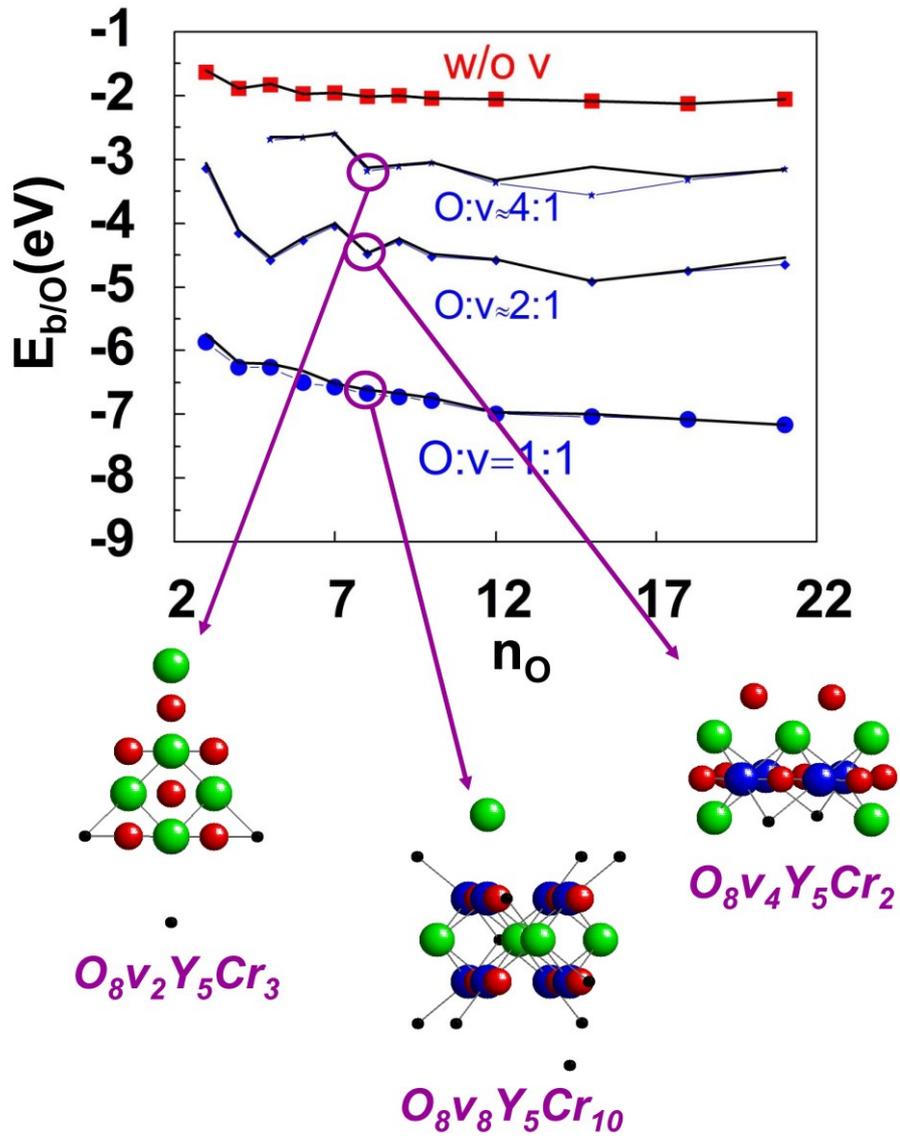

Fig. 5

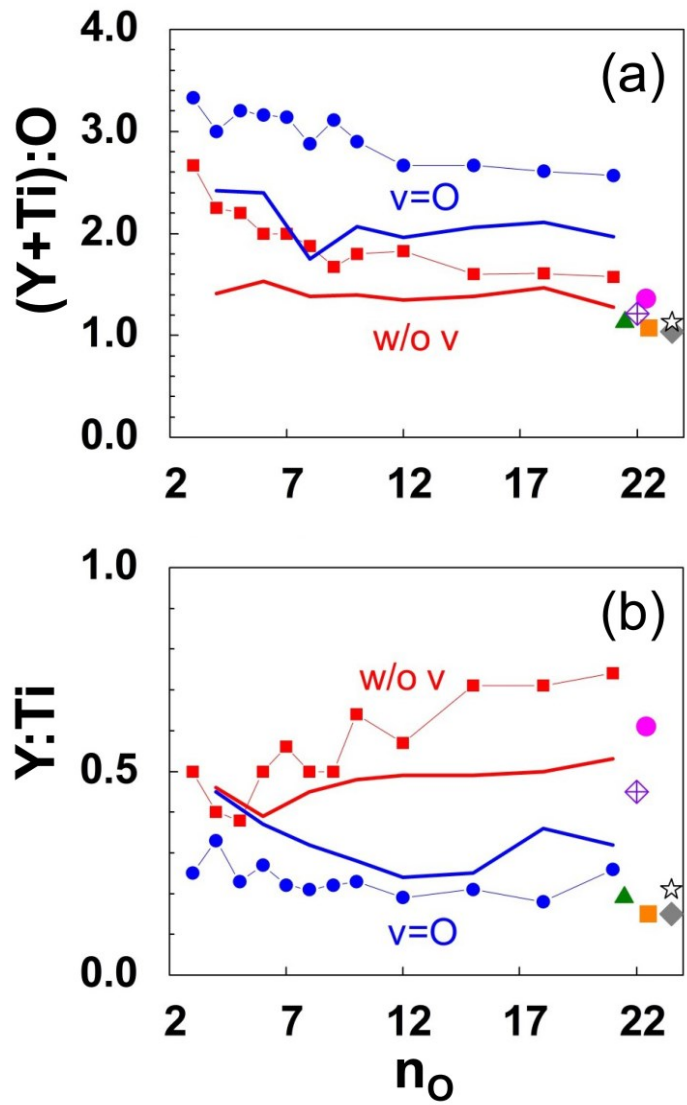

Fig. 6

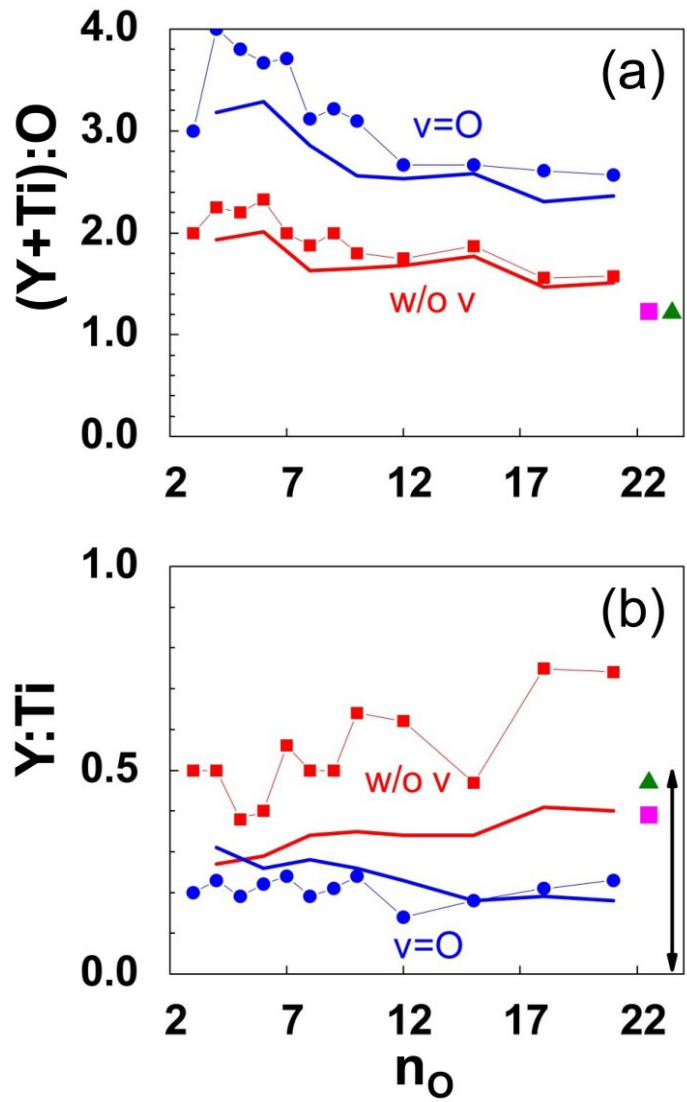

Fig. 7

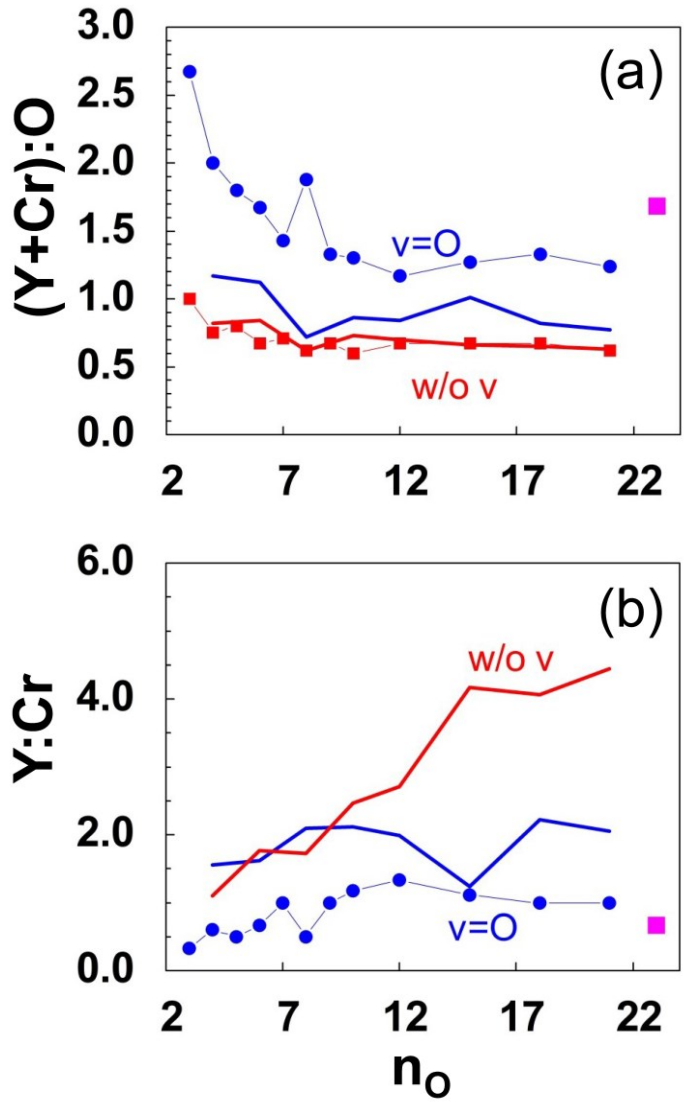

Fig. 8